\documentclass{optica-article}
\journal{opticajournal} 

\articletype{Research Article}

\begin{document}

\title{Programmable Photonic Circuits with Embedded Feedback for Parallel Multi-Wavelength Operations}

\author{Kevin Zelaya\authormark{1,2,3$\dagger$}, Jonathan Friedman\authormark{1,2,3$\dagger$}, and Mohammad-Ali Miri\authormark{1,2,3,*}}

\address{\authormark{1}Department of Physics, Queens College of the City University of New York, Queens, New York 11367, USA\\ \authormark{2}The Graduate Center of the City University of New York, New York, New York 10016, USA\\ \authormark{3}Electrical and Microelectronic Engineering, Rochester Institute of Technology, Rochester, New York 14623, USA }

\email{\authormark{*}ali.miri@rit.edu}

\begin{abstract*}
Linear transformations are cornerstone operations utilized in modern computing, but are computationally expensive on current electronic platforms. Optical computing has been positioned as a new computing solution, promising high speed and energy efficiency by exploiting the available degrees of freedom of light. Although solutions exist in the optical domain, there is a continuous search for compact solutions that properly utilize the limited chip space and exploit various degrees of freedom of light. Here, we introduce and experimentally demonstrate a compact, programmable photonic integrated circuit (PIC) architecture that operates on both spatial and frequency degrees of freedom by leveraging embedded optical feedback loops. This architecture enables universal linear unitary transforms by combining resonators with passive linear mixing layers and tunable active phase layers. The strong dispersion achieved from the resonant loops enables multi-frequency operation and reduces the number of required active layers to achieve universality. This solution reduces the optical port requirements, minimizes power losses, and leverages resonances to enable massive parallel computing in the frequency domain. The fabricated samples are compatible with silicon-on-insulator platforms and operate at single- and dual-frequency modes. The experimental setup demonstrates the ability to perform in situ training in both cases, validating the parallel-computing capabilities of the PICs. This work highlights the potential of feedback-loop PICs for scalable, compact, and energy-efficient linear optical computing.
\end{abstract*}


\section{Introduction}
Photonic technologies are becoming mature and reliable technologies for generating, manipulating, and detecting light at the classical and quantum levels~\cite{madsen2022quantum,maring2024versatile,harris2017quantum}. Photonics has experienced remarkable advancements in recent years, particularly in the visible and near-infrared spectrum. This has driven developments across various domains, including telecommunications and information processing~\cite{zhou2024silicon}, as well as sensor technologies. This includes advancements in various material platforms that support guided light, allowing for precise control, such as silicon-on-silica~\cite{shekhar2024roadmapping}, silicon nitride~\cite{xiang2022silicon}, thin-film lithium niobate~\cite{zhu2021integrated}, and silicon carbide~\cite{yi2022silicon}, among others. These materials platforms serve different purposes, such as the excitation of nonlinear susceptibilities~\cite{phillips2024general}, ultra-fast electro-optical switching~\cite{li2023high}, and low-loss propagation~\cite{ke2022digitally}.

The intersection of photonics and computing is becoming a new frontier in information processing. Optical computing leverages the speed and bandwidth advantages of light over traditional electronic components, promising faster data processing and lower energy consumption. Advances in photonic integrated circuits (PICs) are paving the way for the development of compact and efficient computational devices that could reshape modern computing. Of particular interest are architectures that can optically represent unitary matrices, as they can perform arbitrary optical operations, i.e., universal. Particular realizations of such universal devices are based on meshes of Mach-Zehnder interferometers (MZI) with specific geometries, such as triangular~\cite{reck1994experimental, miller2013self}, rectangular~\cite{clements2016optimal}, diamond~\cite{Shokraneh2020,rahbardar2023addressing}, and hexagonal meshes with protected topological properties~\cite{on2024programmable}. The latter strongly relies on the precision with which MZIs are manufactured, and any defect may render the final device functional. This issue has been recently considered in~\cite{Bandyopadhyay21}, where the authors consider the effects of unitary defects for each MZI. This allows for sequential calibration in meshed architectures, provided certain phases are maintained throughout the calibration process. In the realm of quantum information, photonics provides a platform for developing quantum communication and quantum computing technologies. Photonic systems facilitate the test of quantum effects on a chip-scale device, such as two-photon interference through compact photonic structures~\cite{huang2025massively,estakhri2021tunable}. 

Recent studies have shown promising results in this field, and further research is underway to advance these technologies. Research has been focused on developing devices that can perform all-optical unitary operations packaged in on-chip form-factor architectures~\cite{Bogaerts20a,friedman2025programmable,zelaya2024goldilocks,harris2018linear,miller2013self}. Alternative architectures based on cascading of a fixed intervening operator with diagonal phase shift layers, capable of representing universal unitary matrices, have been reported in the literature \cite{pastor2021arbitrary, Tanomura20, Tanomura23, saygin_robust_2020, Skryabin2021, Markowitz23}. Such an architecture can be realized on-chip by exploiting multimode interference couplers \cite{pastor2021arbitrary} and multiport waveguide couplers interlaced with programmable phase shifters\cite{Tanomura22a,zelaya2024goldilocks}. In particular, recently, we showed that nonuniform photonic lattices of particularly designed coupling coefficients and length to implement a Discrete Fractional Fourier Transform (DFrFT) operation can be utilized as the intervening operation for realizing programmable unitaries through such an interlacing architecture \cite{Markowitz23,Markowitz2023auto}. Further extension to non-unitary linear operations has been demonstrated to be feasible by extending current unitary processors through the singular value decomposition~\cite{miller2013self,giamougiannisUniversalLinearOptics2023} or using novel lossy interlaced structures~\cite{tang2024lower,markowitz2025embedding}.

In this work, we propose a resonant PIC design that exploits unitary multiport couplers and active phase elements in a layered structure. Unlike previous architectures, our design features a fixed number of loops that interconnect output ports to input ports, which reduces the total number of available ports while increasing the density of active components per layer. This approach not only improves component density but also recycles power that would otherwise be lost, circulating it back through the device. The design has been validated through numerical simulations, wave testing, and experimental results. The fabricated sample was produced on a silicon-on-insulator platform, making it suitable for fabrication in open-access silicon foundries.

\section{Results}
\subsection{Design and Fabrication}

\begin{figure}[t!]
    \centering
    \includegraphics[width=0.9\textwidth]{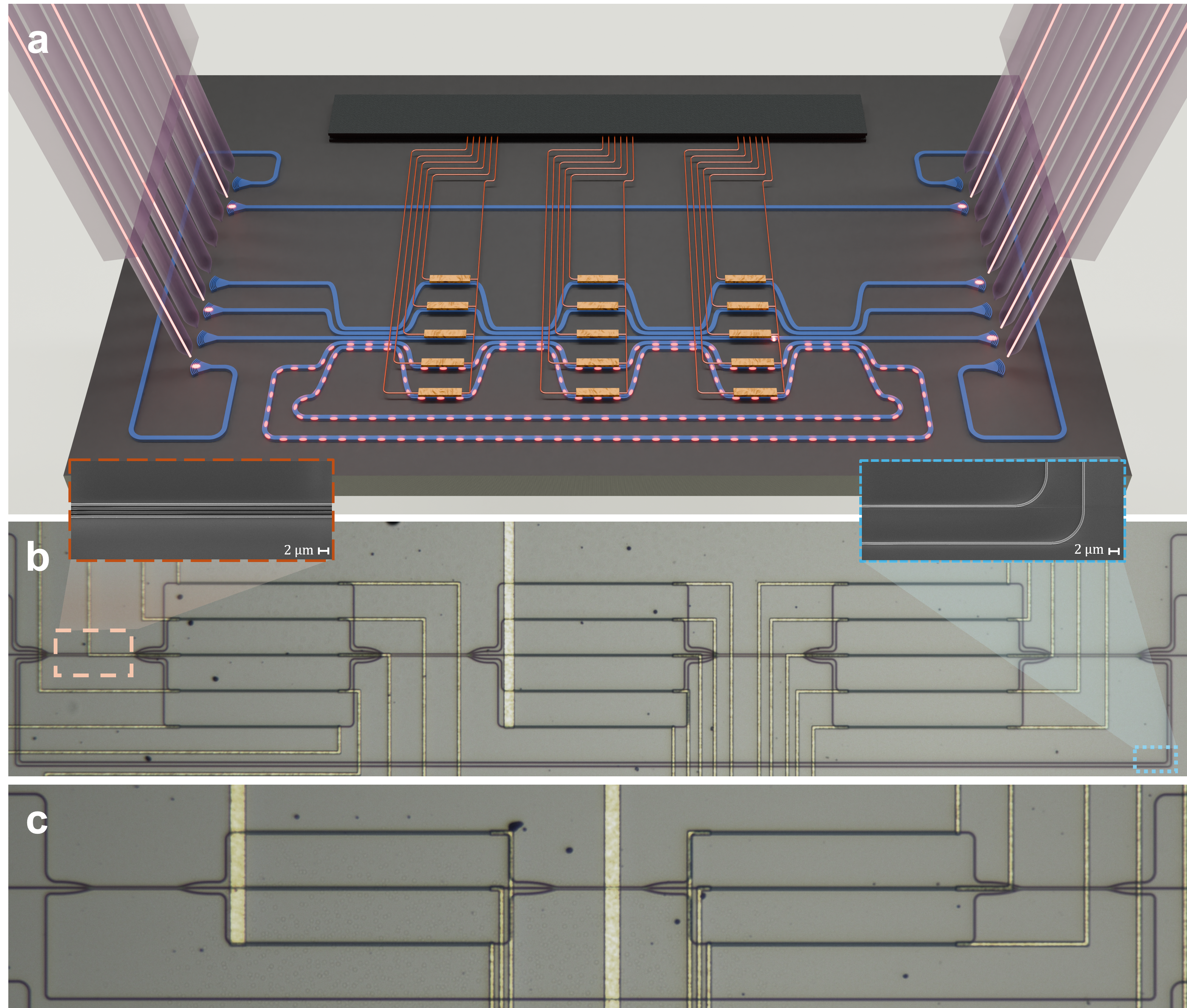}
    \caption{\textbf{Concept and fabricated samples.} \textbf{a} Illustration of one of the programmable looped architectures. This comprises evanescently coupled waveguides used for wave mixing and metal heaters for the phase control required to produce interference. The input and output are coupled to grating couplers to excite and gather the PIC through lensed fibers (left and right side), respectively. \textbf{b} Microscopic image of the fabricated three-port device with three microheater layers and two embedded loops. Insets show the scanning electron microscope (SEM) images of the coupled WGAs (red-dashed area) and the circular bends (blue-dotted area) used to close the loops. \textbf{e} Microscopic image of a second fabricated feedback PIC showcasing a two-port device with two microheater layers and a single loop design.}
    \label{fig:FigX1}
\end{figure}

The present solution for performing linear operations is devised as an interlaced structure composed of passive mixing layers and tunable active phase layers. Unlike previous approaches~\cite{zelaya2024goldilocks,tang2024lower,taguchi2024standalone}, the proposed solution involves a $K$-port device embedded in a $N$-dimensional unitary system, with $K,N\in\mathbb{Z}^{+}$ and $K<N$. The remaining $P=N-K$ ports at the output are looped back into the system input. The loops in the system are deliberately introduced as they serve several purposes. (a) Implementing loops reduces the number of available optical ports, but it allows for the packing of more active elements per layer, ultimately reducing the number of layers $M$ required to achieve universality (see Materials and Methods Section for details). (b) The otherwise leaked power through unused ports~\cite{markowitz2025embedding} is now recirculated throughout the structure, reducing power losses in the system. (c) The appearance of resonances induces a narrow free spectral range (FSR), which opens up the possibility of different linear operations performed across different frequencies. This paper aims to demonstrate the latter points across two photonic integrated circuits (PICs). 

The proposed looped structure is sketched in Fig.~\ref{fig:FigX1}\textbf{a}. This illustrates the waveguides (blue) and the evanescently coupled waveguide arrays (WGAs) used to generate wave mixing across the structure. Micrometal heaters (orange) are used as active elements to tune the structure to the required operation. The designed photonic chip was fabricated through an open-access silicon foundry. The design platform is composed of fully-etched silicon (Si) material deposited on top of a 2 µm-thick SiO${}_{2}$ substrate layer. The silicon layer has a uniform thickness of 220 nm and a variable width, which, for the present design, is kept uniform at 500 nm. The overall structure is buried with a 2.2 µm-thick SiO${}_{2}$ layer. For the active elements, two metallization layers are patterned on top of the cladding for the microheater (high-resistance Ti/W alloy metal) and the routing layer (low-contact resistance TiW/Al bilayer). Then, a thin encapsulation oxide layer is deposited to prevent the wire from breaking and interacting with ambient air. Probing pads connect to the routing layer inside the design area, which are wire-bonded to the printed circuit board (PCB) to enable external electrical connectivity of each phase shifter. 

The chip was packaged with electrical connections. This provides direct electrical access to the metal heaters for programming via an external source measurement unit (SMU), which is controlled through a computer interface. Two V-Groove assembly fiber arrays with a 127 µm pitch and an 8-degree polish angle were mounted on both ends of the chip for light coupling between the input and output grating couplers to the relevant measurement equipment. Several designs are incorporated into the chip, including two programmable looped circuits. Particularly, Fig.~\ref{fig:FigX1}\textbf{b} shows a microscope image of a double-looped design composed of four five-port unitary mixers, achieved through evanescently coupled WGAs (left inset in Fig.~\ref{fig:FigX1}\textbf{b}). This yields a $K=3$ structure with three accessible optical ports at the input and output, featuring 15 active microheaters. Likewise, Fig.~\ref{fig:FigX1}\textbf{c} displays a single-looped structure, where port are used to generate targets in $U(2)$ with two layers of microheaters. In both designs, the WGAs interconnect with other elements in the circuit through cubic B\'ezier bends, designed to minimize mode crosstalk between neighboring waveguides while reducing the power loss due to the bend curvature. This is particularly useful as the spacing among coupled waveguides is in the sub-wavelength scale. In turn, we incorporated circular bends with 10 µm radius to route light through the loops (right inset in Fig.~\ref{fig:FigX1}\textbf{c}).

\subsection{Experimental setup}
\begin{figure}[t!]
    \centering
    \includegraphics[width=0.9\textwidth]{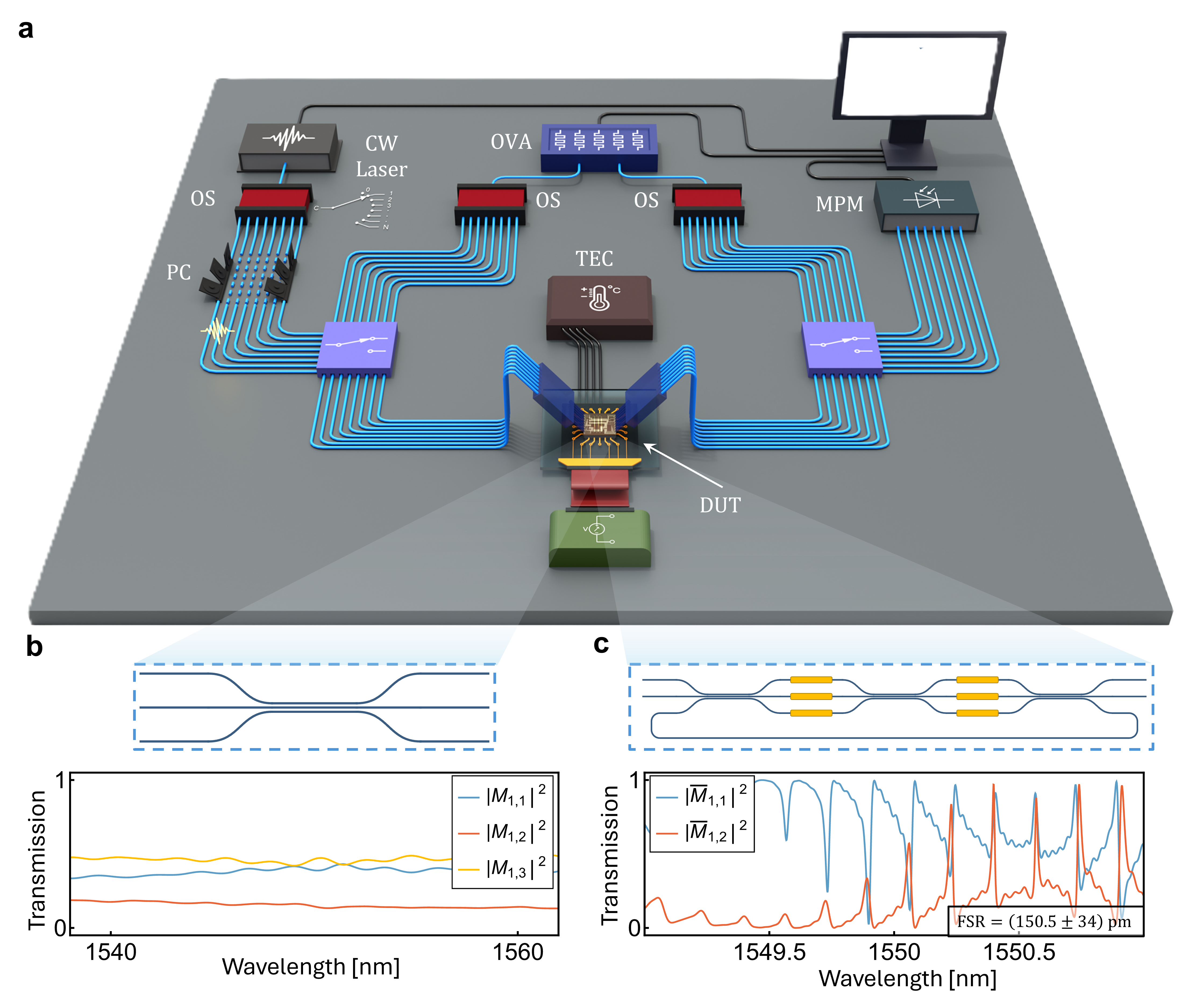}
    \caption{\textbf{Experimental setup and training results.} \textbf{a} Experimental setup used for characterization of the feedback-loop chip. A wavelength-tunable laser source (upper left) interconnects with a programmable 1x8 optical switch (OS), the outputs of which feed into a V-groove fiber array that injects light into the device under test (DUT). A second fiber array couples lights out of the DUT and interconnects with a multiport power meter (MPM) that gathers the optical signal. The DUT is connected to a thermoelectric cooler (TEC) unit for temperature control and to a standard measure unit (SMU) that provides currents to the microheaters in the DUT. A personal computer interfaces with the laser, OS, MPM, SMU, and TEC for control and monitoring.  \textbf{b}-\textbf{c} Dispersion response of a single waveguide guide array ($N=3$) \textbf{b} and a feedback-loop circuit \textbf{c}. The former showcases an almost flat response, whereas the latter depicts the effects of the embedded loops.}
    \label{fig:FigX2}
\end{figure}
The experimental setup implemented to characterize and program the device is sketched in Fig.~\ref{fig:FigX2}\textbf{a}. A wavelength-tunable laser is connected to a multichannel optical switch, which enables the selective excitation of the input grating couplers through an external computer. An 8-channel fiber array links the multichannel optical switch and the grating couplers. We used a polarization controller to optimize the coupling efficiency of the laser light to the grating couplers. The electrical channels are mapped to a PCB and interfaced with a source meter to supply voltage and current to the micro-heater element. The optical output channels are recorded through a multiport power meter. Each measurement consists of selectively exciting one input waveguide at a time and simultaneously capturing the power from all output waveguide arrays after updating the driving current of individual phase shifters. Additionally, a thermoelectric controller (TEC) is incorporated to stabilize the overall temperature of the chip. This is crucial since the optical signal is affected by small fluctuations in temperature, which, in turn, are due to the narrow FSR. 

For the current setup, we operate the circuits in $K=3$ (Fig.~\ref{fig:FigX1}\textbf{b}) and $K=2$ (Fig.~\ref{fig:FigX1}\textbf{c}) at 39 C and 29 C, respectively. This ensures global thermal stability across the chip, even when all heaters are operated at close to maximum power consumption. For this reason, the target temperature is higher on the $K=3$ structure, as more microheaters operate simultaneously in the system.

Individual passive WGAs are included in the fabricated chip and used for independent characterization. Indeed, Fig.~\ref{fig:FigX2}\textbf{b} showcases the coupler utilized in the $K=2$ looped structure in Fig.~\ref{fig:FigX1}\textbf{c}. Such a coupler is designed as a homogeneous WGA with a 350 nm gap between neighboring waveguides and a 110 µm coupling length. These parameters were extracted from mode analysis and FDTD simulations in order to induce a transfer matrix with the density properties required for universality~\cite {zelaya2024goldilocks}. Fig.~\ref{fig:FigX2}\textbf{b} displays the dispersion in the transmission intensities of the passive $3\times 3$ when the uppermost channel is excited  
, showing an almost flat dispersion across the C-band. The presence of the loop in the interlaced structure induces the resonances shown in Fig.~\ref{fig:FigX2}\textbf{c}. The corresponding free spectral range (FSR) yields an average value of around 150.5 pm. This is particularly relevant for the linear operation design at multiple frequencies discussed below.

\subsection{PIC training}
\begin{figure}[t!]
    \centering
    \includegraphics[width=0.9\textwidth]{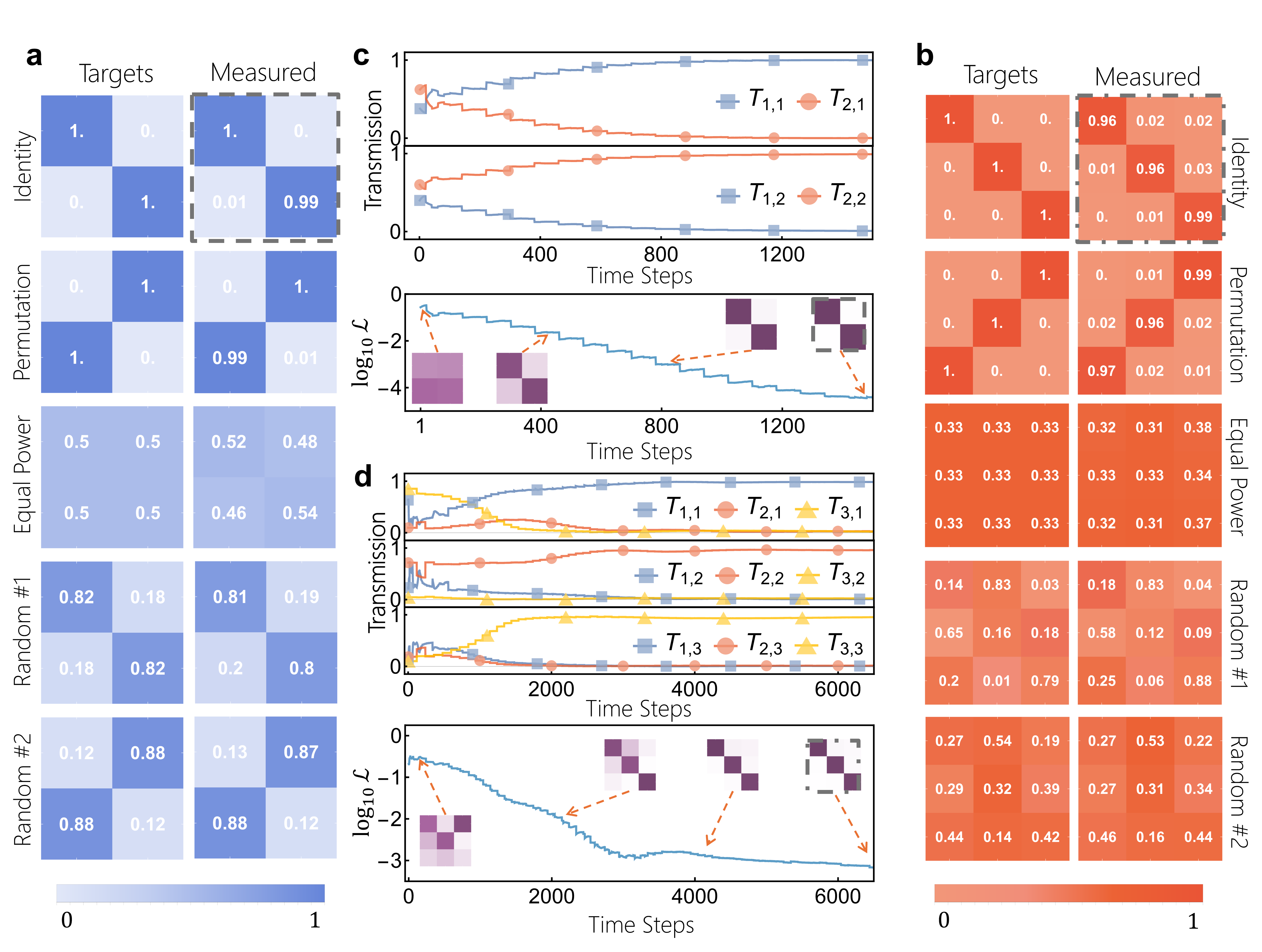}
    \caption{\textbf{Single-frequency training results.} \textbf{a}-\textbf{b} In-situ training and target normalized intensities for the $K=2$ (\textbf{a}) and $K=3$ (\textbf{b}) structures shown in Fig.~\ref{fig:FigX1}\textbf{b}-\textbf{c}. The bar, cross, uniform-power, and random unitaries have been tested in both structures. \textbf{c}-\textbf{d} Training history of the in-situ optimized identity targets and the corresponding figure of merit $\mathcal{L}$. }
    \label{fig:FigX3}
\end{figure}
The looped architecture is operated through \textit{in situ} optimization. Indeed, because of the resonances induced by the loops in the circuit, the device becomes highly sensitive to manufacturing defects and deviations in the currents driving the metal heaters. Thus, to account for all these issues, we run the device by dynamically changing its operation. Henceforth, only intensity measurements and optimizations are performed; however, the current device can be phase-tunable if a suitable phase-measurement device is available. See Method Section for details. Device optimization is performed using gradient-based methods, such as the \textit{adaptive moment estimator} (ADAM), which is widely known in the literature~\cite{kingma2014adam}. The ADAM optimizer is particularly useful because the learning rate is constantly adjusted based on the training session's progress, making it a handy asset as looped structures become highly sensitive to changes. This allows broader changes in the currents when the optimizer begins searching for local minima, then narrows them as it approaches the target.

Be $X(\Phi;\lambda)\in\mathbb{C}^{N_{1}\times N_{2}}$ the measured quantity, with $\Phi$ the set of tunable currents and $\lambda$ the operational wavelength. The figure of merit under consideration is defined through the positive-definite function
\begin{equation}
\mathcal{L}(\Phi;\Lambda) := \frac{\sum_{\lambda\in\Lambda}\Vert X(\Phi;\lambda) - X_{\textnormal{target}}(\lambda) \Vert^2}{N_{1}*N_{2}*N_{\lambda}} , \quad \Lambda:=\{\lambda_{1},\ldots,\lambda_{N_{\lambda}}\},
\label{L}
\end{equation}
where $X_{\textnormal{target}}(\lambda)\in\mathbb{C}^{N_{1}\times N_{2}}$ is the target element at the wavelength $\lambda$, whereas $N_{\lambda}$ is the total number of frequency points sampled from the experiment. In this formulation, we have accounted for the dependence on single or multiple wavelengths through $\Lambda$. The optimizations conducted in this work are aligned with the figure of merit in Eq.~\eqref{L}, where the measured ($X(\Phi;\Lambda)$) and target ($X_{\textnormal{target}}$) quantities are the field intensities (power) of all devices and their corresponding operational setups.

During the experimental test, we consider the $K=2$ and $K=3$ structures shown in Figs.~\ref{fig:FigX1}\textbf{b}-\textbf{c}, in which case, the targets are picked up as elements in the unitary groups $U(2)$ and $U(3)$, respectively. The latter follows from the number of effective ports and the total number of available tuning parameters. The universality of both structures is assessed by optimizing them to reconstruct several unitary targets. This is an expected feature from the analysis presented in the Material and Methods section below, which relies on the over-parameterization of the unitary group in question. 

\subsubsection{Single-frequency operation}
Five targets are selected for each structure, including the bar and cross configurations, uniform-power transmission, and two unitary random targets. The bar-state and cross-state targets are prime examples of sparse matrices (identity and its permutations), whereas the uniform-power matrix is dense. The latter, combined with the random unitaries, spans a broad variety of targets so that the performance of both architectures is not strictly tied to the density or sparsity of the target under test. These targets are encoded in a single operational wavelength $\lambda=1550$ nm ($N_{\lambda}=1$), the results of which are summarized in Fig.~\ref{fig:FigX3}\textbf{a} for $K=2$ and in Fig.~\ref{fig:FigX3}\textbf{b} for $K=3$. 

In all the cases, an in situ training was run using the intensity of unitaries as the target, $X_{\textnormal{target}}(\lambda)=\vert U_t \vert^2$, with $U_t$ one of the selected targets. The current in the heaters is capped at 9.5 mA and transformed through a smoothing function to ensure that $\mathcal{L}$ is a continuous and differentiable (\textit{smooth}) function. Additionally, the training terminates once the figure of merit falls below $5\times 10^{-4}$ or after completing 100 iterations. For example, the training process for the bar-state configuration (identity matrix) is shown in Fig.~\ref{fig:FigX3}\textbf{c} and Fig.~\ref{fig:FigX3}\textbf{d} for both structures. The evolution of the measured power (normalized) and the figure of merit are plotted in terms of the number of time steps. Time steps are internally generated in the multiport power meter unit and correspond to 100 ms of averaged samples. 

\subsubsection{Parallelization via multi-frequency operation}
An alternative functionality can be implemented in the looped structures by encoding information in different wavelengths. Indeed, the selected frequencies cannot be spaced by an integer multiple of the corresponding FSR, nor can they be at the resonant points. This process is tested by training the PICs to achieve two intensity vectors as targets simultaneously, one at 1549.5 nm and another at 1550.5 nm. We thus use $N_{\lambda}=2$ in the figure of merit~\eqref{L} implemented throughout the training. Five combinations of targets are selected, one comprising uniform-power vectors in each vector at both wavelengths. Two targets are used to route light to different output ports at different wavelengths, performing frequency demultiplexing. Two sets of random vectors are encoded at both wavelengths to test cases where no bias is introduced in the target selection. 

\begin{figure}[t!]
    \centering
    \includegraphics[width=0.7\textwidth]{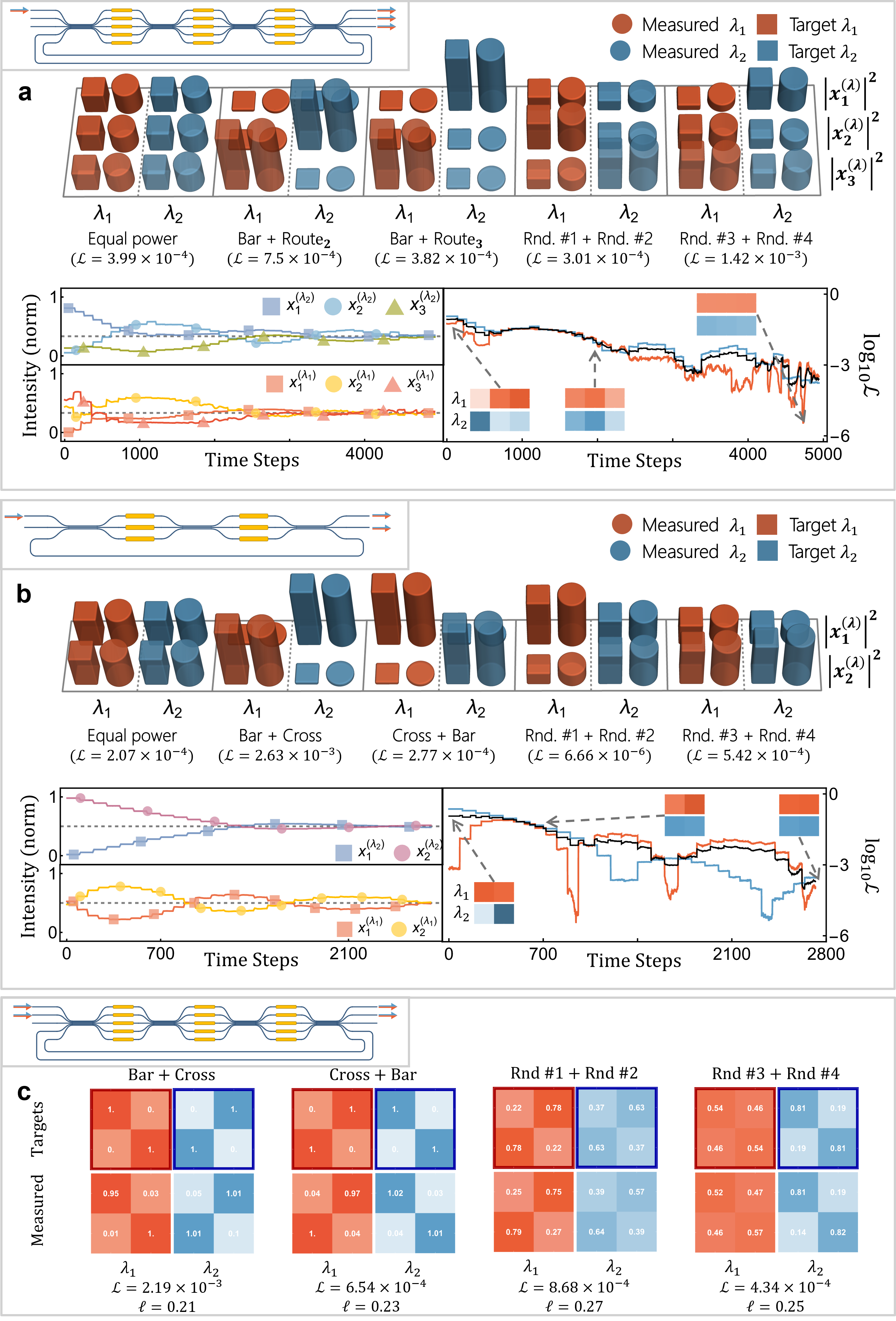}
    \caption{\textbf{Parallel linear transforms.} \textbf{a}-\textbf{b} Parallel and arbitrary spatial photonic state generation from a single excitation at the upper-most port for the $K=2$ (\textbf{a}) and $K=3$ (\textbf{b}) structures at 1549.5 nm and 1550.5 nm. The lower-left and lower-right panels show the training history and figure of merit history, respectively, for the uniform photonic state generation. \textbf{c} Parallel matrix-vector operations by simultaneously embedding two $2\times 2$ unitaries in the $K=3$ structure at 1549.5 nm and 1550.5 nm wavelengths.}
    \label{fig:FigX4}
\end{figure}

In all cases, only the uppermost input port is excited, while the optical power is measured across all output ports. This renders a $K$-dimensional vector per wavelength. The experimental results are summarized in Fig.~\ref{fig:FigX4}\textbf{a} and Fig.~\ref{fig:FigX4}\textbf{b} for the $K=2$ and $K=3$ structures, respectively. For illustration, bar plots of both the targets (bars) and trained measured intensities (cylinders) are presented at 1549.5 nm (red) and 1550.5 nm (blue) side-by-side for direct comparison. For completeness, the corresponding figure of merit is shown to assess performance numerically. The prescribed tolerance in the figure of merit is achieved in most cases, except for a few samples. Still, the reconstructed vectors do not deviate significantly from the corresponding target. This is primarily due to the low tolerance error used during the training, which forces the PIC to steer as close as possible to the target. As a result, even when $\mathcal{L}$ is an order of magnitude greater, the outcomes remain within acceptable limits. 

Finally, two unitaries can be encoded simultaneously into one of the structures. However, the $K=2$ structure includes six tunable parameters, which is less than the minimum number of required parameters $2K^2$. The same holds for the $K=3$ structure. To overcome this issue, we can embed lower-dimensional unitaries into the $K=3$ structure, an approach that effectively increases the number of parameters by introducing losses~\cite{markowitz2025embedding}. This is achieved by encoding two $2\times 2$ unitaries in the $K=3$ structure by exciting two input ports and measuring the intensity from two output ports. Since an output port is left open, there is a power leakage $\ell=1-g^{-1}$, with $g$ a gain factor required to match the rescaled embed unitaries with the desired target. The figure of merit used during the training follows from Eq.~\eqref{L} by using $X(\Phi;\lambda)\rightarrow g \overline{X}(\Phi;\lambda)$, where $\overline{X}(\Phi;\lambda)$ is a $2\times 2$ section of the $K=3$ structure, usually the upper-left section of the transmission matrix. Furthermore, $g\in[1,2]$ is a trainable parameter whose upper bound limits the maximum leakage allowed during the training. We evaluated four distinct combinations of two unitary targets at wavelengths of 1549.5 nm and 1550.5 nm, and the findings are summarized in Fig.~\ref{fig:FigX4}\textbf{c}. The worst-performing case yields a figure of merit an order of magnitude higher than the prescribed tolerance, which, just like in the parallel vector case, still achieves acceptable tolerance in reconstructing the intended target. However, this approach induces loss factors $\ell=21\% \sim 27\%$ across the tested configurations.


\section{Discussion}
The proposed design and fabricated photonic PICs have been thoroughly characterized and validated across various experimental conditions. Indeed, we have demonstrated the capability to experimentally encode unitaries with high accuracy within feedback loop architectures, effectively reducing the number of active layers required. Conventional interlaced architectures utilize a significant number of layers, such as the MMI-based design, which requires $6N+1$ layers~\cite{pastor2021arbitrary}, or the WGA-based approach with $N+1$ layers~\cite{zelaya2024goldilocks,taguchi2024standalone}. Our architecture enables the in situ training and implementation of $K=2$ and $K=3$ unitaries using precisely $K$ layers. The latter is achieved by integrating feedback loops, which allows for the incorporation of additional active components within each layer, thus mitigating the need for extra layers to achieve full parameterization of the desired unitary transformations. Most targeted unitaries have consistently yielded results within the specified accuracy threshold. Although a few targets reproduced results below the prescribed tolerance, the discrepancies between the target and reconstructed unitaries proved negligible. 

Two-frequency parallelization has been experimentally shown to be effectively implemented in linear operations, such as multi-frequency photonic state preparation and matrix-vector multiplication. In both instances, device performance was evaluated at wavelengths of 1549.5 nm and 1550.5 nm, as initially designed. Further parallelization can be achieved by designing the circuit effectively with the appropriate number of loops and phase shifters.

It is worth noting that the transfer matrix of the looped structure in Eq.~\ref{M_tilde} (see Materials and Methods section), implicitly involves a term of the form $(\mathbb{I}-\mathbb{M}_{4}^{\gamma})^{-1}$. This can be further exploited to perform matrix inversion operations if the device is designed and operated correctly. In the current design, such an operation is not possible as the individual sub-matrices $\mathbb{M}_{j}$ are not accessible. Nevertheless, a modified device can achieve the matrix inversion task. Results in this regard will be reported elsewhere.





\section{Materials and Methods}

\begin{figure}[t!]
    \centering
    \includegraphics[width=0.95\textwidth]{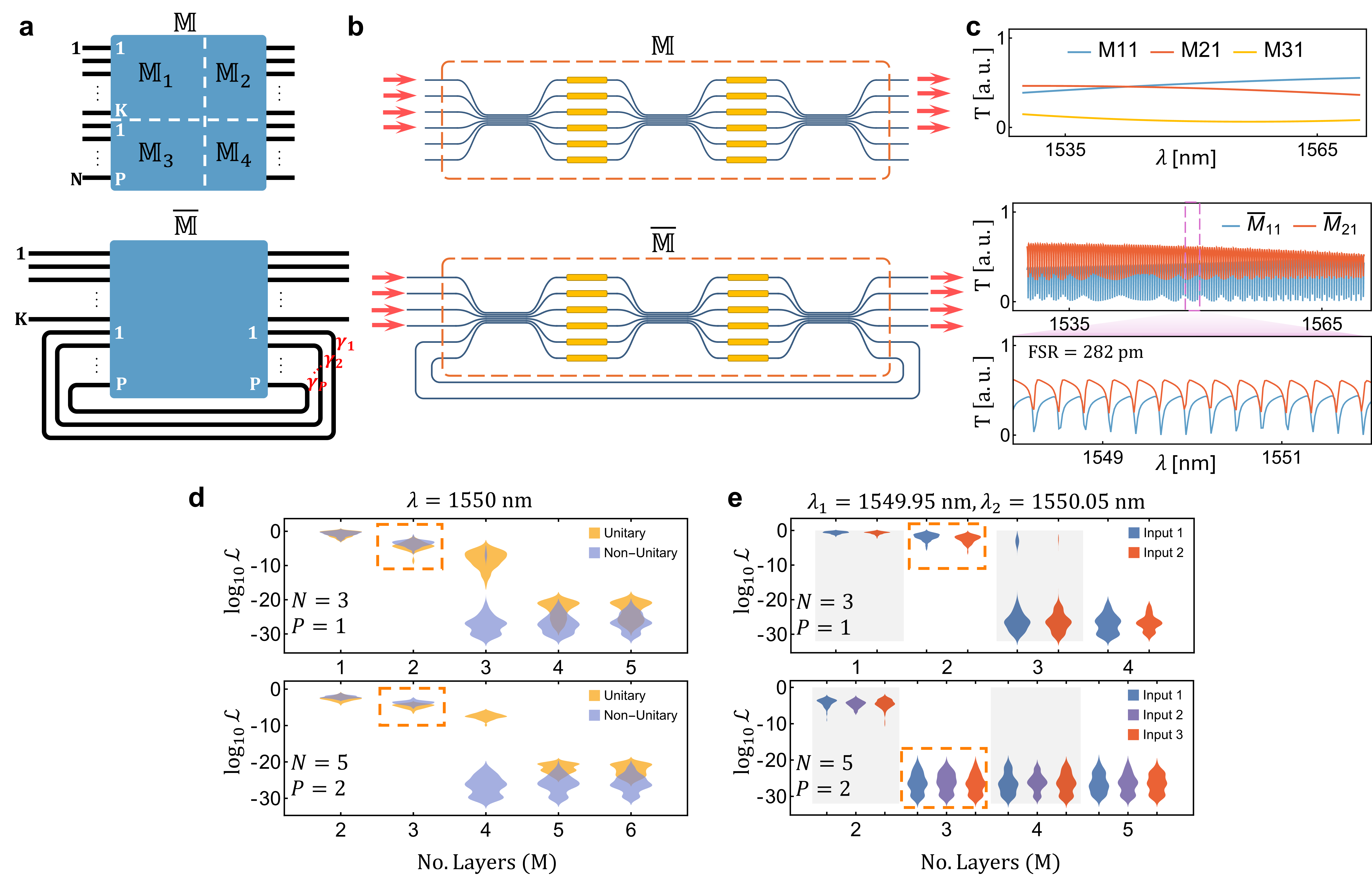}
    \caption{\textbf{Theoretical model and predictions of the looped structures.} \textbf{a} Schematic depiction of the transmission matrix of a straight structure (upper panel) and its decomposition (lower panel) into the smaller-dimensional matrices $\mathbb{M}_{1,2,3,4}$. The latter builds up the looped structure $\overline{\mathbb{M}}$. \textbf{b} Examples of the straight and feedback-loop structures for $N=3$ and $P=1$ (one loop). \textbf{c} Predicted wavelength response for the straight and resonant structures. This indicates an FSR of approximately 282 pm. \textbf{d} Single-wavelength numerical optimizations of the feedback-loop structure using the dispersive model introduced in the text for $\{N=3,P=1\}$ and $\{N=5, P=2\}$, and several numbers of layers. In this test, 100 random unitary and non-unitary matrices are randomly generated and used as targets. \textbf{e} Numerical optimization for random vectors as targets in the dual wavelength operation at 1549.5 nm and 1550.5 nm. Here, 100 random normalized vectors for $\{N=3,P=1\}$ and $\{N=5, P=2\}$, and various number of layers. }
    \label{fig:FigX5}
\end{figure}

Let us consider a linear operator $\mathbb{M}\in\mathbb{C}^{N\times N}$ that characterizes the wave evolution of an optical component. Although most practical cases relate to unitary evolutions $\mathcal{U}\in U(N)$, the theory applies to any general linear transform. In a previous work~\cite{markowitz2025embedding}, it was shown that embedding smaller dimensional matrices into larger unitary and non-unitary matrices is possible, rendering flexible PIC designs. Despite the unavoidable power losses induced by open ports, this design method remains robust. Thus, to avoid such losses, we consider the case in which the open ports are used to recirculate the otherwise lost light by recirculating it back into the circuit. Indeed, this process induces resonances in the circuit that ultimately reduce the free-spectral range of the device.


The transmission matrix $\overline{\mathbb{M}} \in \mathbb{C}^{K \times K}$ for the feedback loop structure integrated into an \(N\)-dimensional device, characterized by $P = N - K$ loops, is derived through an iterative analysis of wave propagation within the architecture for a chosen excitation profile. The latter is determined from the transmission matrix $\mathbb{M}$ of the straight (non-looped) device (upper panel in Fig.~\ref{fig:FigX5}\textbf{a}). The output of the corresponding looped structure (lower panel in Fig.~\ref{fig:FigX5}\textbf{a}) is achieved in the \textit{steady-state} regime when a sufficient number of iterations are performed; that is, in the asymptotic limit. To this end, the transmission matrix $\mathbb{M}\in\mathbb{C}^{N\times N}$ is conveniently decomposed into the rectangular and lower-dimensional matrices
\begin{equation}
\mathbb{M}\equiv\begin{pmatrix}
\mathbb{M}_{1} & \mathbb{M}_{2} \\ 
\mathbb{M}_{3} & \mathbb{M}_{4}
\end{pmatrix}
\label{eq_M}
\end{equation}
where $\mathbb{M}_{1}\in\mathbb{C}^{(N-P)\times(N-P)}$, $\mathbb{M}_{2}\in\mathbb{C}^{(N-P)\times P}$, $\mathbb{M}_{3}\in\mathbb{C}^{P\times (N-P)}$, and $\mathbb{M}_{4}\in\mathbb{C}^{P\times P}$. Furthermore, each loop might induce losses and additional phases into the PIC, which we account for by the factors $\gamma_{q}\in\mathbb{C}$, with $\vert\gamma_{q}\vert\leq 1$ and $q\in\{1,\ldots,P\}$. 

Now, let us consider $\boldsymbol{x}\in\mathbb{C}^{N-P}$ as the input vector fed into the looped architecture. The steady-state solution $\overline{\boldsymbol{x}}\in\mathbb{C}^{N-P}$ in the device output is achieved after several iterations through the device. The latter renders the output state $\overline{\boldsymbol{x}}=\overline{\mathbb{M}}\boldsymbol{x}$, where
\begin{equation}
\overline{M}:= \mathbb{M}_{1} + \mathbb{M}_{2}(\mathbb{I}_{P\times P}-\mathbb{M}_{4}^{\gamma})^{-1}\mathbb{M}_{3}^{\gamma} , \quad \mathbb{M}_{3,4}^{\gamma}:=\mathbb{D}_{\gamma}\mathbb{M}_{3,4} ,
\label{M_tilde}
\end{equation}
is the transmission matrix of the feedback loop device, $\mathbb{D}_{\gamma}:=\textnormal{diag}\{\gamma_{1},\ldots,\gamma_{P}\}$ is a diagonal matrix with the looped factors $\gamma_{p}$ in its diagonal, and $\mathbb{I}_{P\times P}$ the corresponding identity matrix. 

Particular realizations are achievable by considering unitary transmission matrices $\mathbb{M}=U(\Phi)\in U(N)$ for the straight structure, such as interlaced circuits~\cite{pastor2021arbitrary,Markowitz23,markowitz2024learning,zelaya2024goldilocks,tanomura2022scalable,taguchi2024standalone} and meshes of MZIs~\cite{reck1994experimental,clements2016optimal}. Throughout the manuscript, we chose interlaced structures as the feed-forward device $\mathbb{M}$, such as the one depicted in the upper panel of Fig.~\ref{fig:FigX5}\textbf{b}. Such a design is based on waveguide arrays implemented as passive elements and phase shifters for the active components. These systems are known to exhibit universality when $M=N+1$ layers are used~\cite{zelaya2024goldilocks}, covering the whole unitary group $U(N)$. The transmission $\overline{\mathbb{M}}$ of the loop-feedback design, shown in the lower panel of Fig.~\ref{fig:FigX5}\textbf{b}, achieves the same universality by reducing the device size in one direction while increasing it in the other, making more efficient use of the chip area. 


The wavelength dependence response of the straight ($\mathbb{M}$) and looped ($\overline{\mathbb{M}}$) structures can be incorporated in the theoretical description. To this end, we run EM simulations to determine the behavior of coupled waveguides for wavelengths in the C-band. These results show that the coupling parameter $\kappa$ has a linear response on $\lambda$ and an exponential response on the gap $s$, which can be expressed as 
\begin{equation}
\kappa(s;\lambda)=(a_{0}+a_{1}\lambda)e^{-\ell s}, 
\label{eq:dispersive_model}
\end{equation}
where the parameters $a_{0}$, $a_{1}$, and $\ell$ are fitted from simulation results. In turn, each loop induces a wavelength behavior of the form $\gamma=\vert \gamma\vert \exp\left\{-i\frac{2\pi}{\lambda}n_{\textnormal{mode}}(\lambda)L \right\}$. Here, $L$ is the total optical path covered by the loop and $n_{\textnormal{mode}}(\lambda)$ is the mode index of the fundamental TE mode as a function of $\lambda$, which is linear in $\lambda$. 

The corresponding wavelength dispersions in the transmission intensities of the straight structure with $N=3$ and $M=2$ layers are computed by randomly fixing the phases. The resulting dispersion is shown in Fig.~\ref{fig:FigX5}\textbf{c} (upper panel), showcasing the expected flat response. In turn, the response of the feedback loop structure is computed from Eq.~\eqref{M_tilde} by adding a single loop of length 2.12 mm, the length used in the fabricated sample. The resulting dispersion and FSR are displayed in Fig.~\ref{fig:FigX5}\textbf{c} (lower panel).

We developed theoretical models for the two fabricated samples: one with parameters $\{ N=3, P=1 $ and $ N=5, P=2\}$. In our performance assessment, we consider various numbers of layers, $M$. Additionally, we randomly generated 100 Haar random unitary matrices, as well as 100 non-unitary matrices using singular value decomposition with singular values spread randomly in the interval $(0, 1)$. The optimization for a single wavelength (1550 nm) is performed using the figure of merit from Equation \eqref{L} for several different values of \(M\), resulting in \(NM\) trainable parameters. The results for both unitary and non-unitary targets are presented in Fig. \ref{fig:FigX5}\textbf{d}. Although using \(M=2\) layers for \(K=2\) did not yield the best performance overall, it struck an optimal balance between accuracy and size for unitary targets. This observation was confirmed experimentally in the Results section. A similar conclusion applies to the device with \(N=5\). To performance of our fabricated sample for non-unitary targets operations can be improved by adding one additional layer to the circuit.


On the other hand, the parallel computing capabilities of the device can be theoretically examined using the dispersive coupling $\kappa$ from Eq.~\eqref{eq:dispersive_model}. Although this approach offers limited dispersion on the passive couplers, combining multiple couplers enhances the wavelength dependence, as demonstrated theoretically and experimentally in Ref.~\cite{friedman2025programmable}. The interplay between these couplers and the feedback loops ultimately enables the dispersion utilized for encoding parallel operations.

In this work, we evaluate the theoretical dual-wavelength operation of the device by generating two sets of randomly normalized vectors, $ \{ \textbf{x}_{p}^{(\lambda_{j})} \}_{p=1}^{100} $, where $ j\in\{1,2\} $. Since the targets are vectors, the device is excited through one of its inputs, which reduces the transmission matrix to a vector. This allows us to implement the optimization problem in a manner compatible with the figure of merit in Eq.~\eqref{L}. Since we investigate the theoretical models related to two fabricated samples, we choose the parameters $\{ N=3, P=1$ and $\{ N=5, P=2 \}$ along with different numbers of layers $M$ to assess performance. Numerical results are summarized in Fig.~\ref{fig:FigX5}\textbf{e} for different excitation ports and for the dual wavelengths of 1549.95 nm and 1550.05 nm. The dashed boxes denote the cases associated with the fabricated samples.

\begin{backmatter}

\bmsection{Author Contributions} 
K.Z. developed the theoretical model, conducted numerical and electromagnetic simulations, and wrote the initial draft of the manuscript. J.F. operated the experimental setup, coupled the optical chips, and automated the measurement process. K.Z. and J.F. prepared the PIC layout and performed the in-situ optimization. M.A.M. conceived the original idea and supervised the project. All authors contributed to the revision of the manuscript.

\bmsection{Funding}
This project is supported by the U.S. Air Force Office of Scientific Research (AFOSR) Awards\# FA9550-22-1-0189 and FA9550-25-1-0200, the Defense University Research Instrumentation Program (DURIP) Award\# FA9550-23-1-0539, and the National Science Foundation Award\# CNS-2329021.

\bmsection{Conflicts of Interest}
The authors declare that there is no conflict of interest regarding the publication of this article.

\bmsection{Data Availability}
The datasets generated and analyzed during the current study are available from the corresponding 
author on reasonable request.

\end{backmatter}

\bibliography{sample}

\begin{thebibliography}{10}
\newcommand{\enquote}[1]{``#1''}

\bibitem{madsen2022quantum}
L.~S. Madsen, F.~Laudenbach, M.~F. Askarani, \emph{et~al.}, \enquote{Quantum
  computational advantage with a programmable photonic processor,}
  {\protect\JournalTitle{Nature}} \textbf{606}, 75--81 (2022).

\bibitem{maring2024versatile}
N.~Maring, A.~Fyrillas, M.~Pont, \emph{et~al.}, \enquote{A versatile
  single-photon-based quantum computing platform,}
  {\protect\JournalTitle{Nature Photonics}} \textbf{18}, 603--609 (2024).

\bibitem{harris2017quantum}
N.~C. Harris, G.~R. Steinbrecher, M.~Prabhu, \emph{et~al.}, \enquote{Quantum
  transport simulations in a programmable nanophotonic processor,}
  {\protect\JournalTitle{Nature Photonics}} \textbf{11}, 447--452 (2017).

\bibitem{zhou2024silicon}
X.~Zhou, D.~Yi, D.~W.~U. Chan, and H.~K. Tsang, \enquote{Silicon photonics for
  high-speed communications and photonic signal processing,}
  {\protect\JournalTitle{npj Nanophotonics}} \textbf{1}, 27 (2024).

\bibitem{shekhar2024roadmapping}
S.~Shekhar, W.~Bogaerts, L.~Chrostowski, \emph{et~al.}, \enquote{Roadmapping
  the next generation of silicon photonics,} {\protect\JournalTitle{Nature
  Communications}} \textbf{15}, 751 (2024).

\bibitem{xiang2022silicon}
C.~Xiang, W.~Jin, and J.~E. Bowers, \enquote{Silicon nitride passive and active
  photonic integrated circuits: trends and prospects,}
  {\protect\JournalTitle{Photonics research}} \textbf{10}, A82--A96 (2022).

\bibitem{zhu2021integrated}
D.~Zhu, L.~Shao, M.~Yu, \emph{et~al.}, \enquote{Integrated photonics on
  thin-film lithium niobate,} {\protect\JournalTitle{Advances in Optics and
  Photonics}} \textbf{13}, 242--352 (2021).

\bibitem{yi2022silicon}
A.~Yi, C.~Wang, L.~Zhou, \emph{et~al.}, \enquote{Silicon carbide for integrated
  photonics,} {\protect\JournalTitle{Applied Physics Reviews}} \textbf{9}
  (2022).

\bibitem{phillips2024general}
C.~R. Phillips, M.~Jankowski, N.~Flemens, and M.~Fejer, \enquote{General
  framework for ultrafast nonlinear photonics: unifying single and
  multi-envelope treatments,} {\protect\JournalTitle{Optics Express}}
  \textbf{32}, 8284--8307 (2024).

\bibitem{li2023high}
Y.~Li, T.~Lan, D.~Yang, \emph{et~al.}, \enquote{High-performance mach--zehnder
  modulator based on thin-film lithium niobate with low voltage-length
  product,} {\protect\JournalTitle{ACS omega}} \textbf{8}, 9644--9651 (2023).

\bibitem{ke2022digitally}
W.~Ke, Y.~Lin, M.~He, \emph{et~al.}, \enquote{Digitally tunable optical delay
  line based on thin-film lithium niobate featuring high switching speed and
  low optical loss,} {\protect\JournalTitle{Photonics Research}} \textbf{10},
  2575--2583 (2022).

\bibitem{reck1994experimental}
M.~Reck, A.~Zeilinger, H.~J. Bernstein, and P.~Bertani, \enquote{Experimental
  realization of any discrete unitary operator,}
  {\protect\JournalTitle{Physical review letters}} \textbf{73}, 58 (1994).

\bibitem{miller2013self}
D.~A. Miller, \enquote{Self-configuring universal linear optical component,}
  {\protect\JournalTitle{Photonics Research}} \textbf{1}, 1--15 (2013).

\bibitem{clements2016optimal}
W.~R. Clements, P.~C. Humphreys, B.~J. Metcalf, \emph{et~al.}, \enquote{Optimal
  design for universal multiport interferometers,}
  {\protect\JournalTitle{Optica}} \textbf{3}, 1460--1465 (2016).

\bibitem{Shokraneh2020}
F.~Shokraneh, S.~Geoffroy-Gagnon, and O.~Liboiron-Ladouceur, \enquote{The
  diamond mesh, a phase-error-and loss-tolerant field-programmable mzi-based
  optical processor for optical neural networks,} {\protect\JournalTitle{Optics
  Express}} \textbf{28}, 23495--23508 (2020).

\bibitem{rahbardar2023addressing}
K.~Rahbardar~Mojaver, B.~Zhao, E.~Leung, \emph{et~al.}, \enquote{Addressing the
  programming challenges of practical interferometric mesh based optical
  processors,} {\protect\JournalTitle{Optics Express}} \textbf{31},
  23851--23866 (2023).

\bibitem{on2024programmable}
M.~B. On, F.~Ashtiani, D.~Sanchez-Jacome, \emph{et~al.}, \enquote{Programmable
  integrated photonics for topological hamiltonians,}
  {\protect\JournalTitle{Nature Communications}} \textbf{15}, 629 (2024).

\bibitem{Bandyopadhyay21}
S.~Bandyopadhyay, R.~Hamerly, and D.~Englund, \enquote{Hardware error
  correction for programmable photonics,} {\protect\JournalTitle{Optica}}
  \textbf{8}, 1247--1255 (2021).

\bibitem{huang2025massively}
L.~Huang, W.~Wang, F.~Wang, \emph{et~al.}, \enquote{Massively parallel
  hong-ou-mandel interference based on independent soliton microcombs,}
  {\protect\JournalTitle{Science Advances}} \textbf{11}, eadq8982 (2025).

\bibitem{estakhri2021tunable}
N.~M. Estakhri and T.~B. Norris, \enquote{Tunable quantum two-photon
  interference with reconfigurable metasurfaces using phase-change materials,}
  {\protect\JournalTitle{Optics Express}} \textbf{29}, 14245--14259 (2021).

\bibitem{Bogaerts20a}
W.~Bogaerts, D.~P{\'e}rez, J.~Capmany, \emph{et~al.}, \enquote{Programmable
  photonic circuits,} {\protect\JournalTitle{Nature}} \textbf{586}, 207--216
  (2020).

\bibitem{friedman2025programmable}
J.~Friedman, K.~Zelaya, M.~Honari-Latifpour, and M.-A. Miri,
  \enquote{Programmable space-frequency linear transformations in photonic
  interlacing architectures,} {\protect\JournalTitle{Scientific Reports}}
  \textbf{15}, 35173 (2025).

\bibitem{zelaya2024goldilocks}
K.~Zelaya, M.~Markowitz, and M.-A. Miri, \enquote{The goldilocks principle of
  learning unitaries by interlacing fixed operators with programmable phase
  shifters on a photonic chip,} {\protect\JournalTitle{Scientific Reports}}
  \textbf{14}, 10950 (2024).

\bibitem{harris2018linear}
N.~C. Harris, J.~Carolan, D.~Bunandar, \emph{et~al.}, \enquote{Linear
  programmable nanophotonic processors,} {\protect\JournalTitle{Optica}}
  \textbf{5}, 1623--1631 (2018).

\bibitem{pastor2021arbitrary}
V.~L. Pastor, J.~Lundeen, and F.~Marquardt, \enquote{Arbitrary optical wave
  evolution with {F}ourier transforms and phase masks,}
  {\protect\JournalTitle{Optics Express}} \textbf{29}, 38441--38450 (2021).

\bibitem{Tanomura20}
R.~Tanomura, R.~Tang, S.~Ghosh, \emph{et~al.}, \enquote{Robust integrated
  optical unitary converter using multiport directional couplers,}
  {\protect\JournalTitle{Journal of Lightwave Technology}} \textbf{38}, 60--66
  (2020).

\bibitem{Tanomura23}
R.~Tanomura, T.~Tanemura, and Y.~Nakano, \enquote{Multi-wavelength
  dual-polarization optical unitary processor using integrated multi-plane
  light converter,} {\protect\JournalTitle{Japanese Journal of Applied
  Physics}} p. SC1029 (2023).

\bibitem{saygin_robust_2020}
M.~Y. Saygin, I.~V. Kondratyev, I.~V. Dyakonov, \emph{et~al.}, \enquote{Robust
  {Architecture} for {Programmable} {Universal} {Unitaries},}
  {\protect\JournalTitle{Physical Review Letters}} \textbf{124}, 010501 (2020).

\bibitem{Skryabin2021}
N.~Skryabin, I.~Dyakonov, M.~Y. Saygin, and S.~Kulik,
  \enquote{Waveguide-lattice-based architecture for multichannel optical
  transformations,} {\protect\JournalTitle{Optics Express}} \textbf{29},
  26058--26067 (2021).

\bibitem{Markowitz23}
M.~Markowitz and M.-A. Miri, \enquote{Universal unitary photonic circuits by
  interlacing discrete fractional fourier transform and phase modulation,}
  (2023).

\bibitem{Tanomura22a}
R.~Tanomura, R.~Tang, T.~Umezaki, \emph{et~al.}, \enquote{Scalable and robust
  photonic integrated unitary converter based on multiplane light conversion,}
  {\protect\JournalTitle{Physical Review Applied}} \textbf{17}, 024071 (2022).

\bibitem{Markowitz2023auto}
M.~Markowitz, K.~Zelaya, and M.-A. Miri, \enquote{Auto-calibrating universal
  programmable photonic circuits: hardware error-correction and defect
  resilience,} {\protect\JournalTitle{Optics Express}} \textbf{31},
  37673--37682 (2023).

\bibitem{giamougiannisUniversalLinearOptics2023}
G.~Giamougiannis, A.~Tsakyridis, M.~{Moralis-Pegios}, \emph{et~al.},
  \enquote{Universal {{Linear Optics Revisited}}: {{New Perspectives}} for
  {{Neuromorphic Computing With Silicon Photonics}},}
  {\protect\JournalTitle{IEEE Journal of Selected Topics in Quantum
  Electronics}} \textbf{29}, 1--16 (2023).

\bibitem{tang2024lower}
R.~Tang, R.~Tanomura, T.~Tanemura, and Y.~Nakano, \enquote{Lower-depth
  programmable linear optical processors,} {\protect\JournalTitle{Physical
  Review Applied}} \textbf{21}, 014054 (2024).

\bibitem{markowitz2025embedding}
M.~Markowitz, K.~Zelaya, and M.-A. Miri, \enquote{Embedding matrices in
  programmable photonic networks with flexible depth and width,}
  {\protect\JournalTitle{Optics Letters}} \textbf{50}, 2318--2321 (2025).

\bibitem{taguchi2024standalone}
Y.~Taguchi and Y.~Ozeki, \enquote{Standalone gradient measurement of matrix
  norm for programmable unitary converters,} {\protect\JournalTitle{JOSA B}}
  \textbf{41}, 1425--1431 (2024).

\bibitem{kingma2014adam}
D.~P. Kingma and J.~Ba, \enquote{Adam: A method for stochastic optimization,}
  {\protect\JournalTitle{arXiv preprint arXiv:1412.6980}}  (2014).

\bibitem{markowitz2024learning}
M.~Markowitz, K.~Zelaya, and M.-A. Miri, \enquote{Learning arbitrary complex
  matrices by interlacing amplitude and phase masks with fixed unitary
  operations,} {\protect\JournalTitle{Physical Review A}} \textbf{110}, 033501
  (2024).

\bibitem{tanomura2022scalable}
R.~Tanomura, R.~Tang, T.~Umezaki, \emph{et~al.}, \enquote{Scalable and robust
  photonic integrated unitary converter based on multiplane light conversion,}
  {\protect\JournalTitle{Physical Review Applied}} \textbf{17}, 024071 (2022).

\end{thebibliography}

\end{document}